\documentclass[conference]{IEEEtran}
\IEEEoverridecommandlockouts
\usepackage{cite}
\usepackage{amsmath,amssymb,amsfonts}
\usepackage{algorithmic}
\usepackage{graphicx}
\usepackage{textcomp}
\usepackage{xcolor}
\usepackage{float}
\usepackage{makecell}
\usepackage{dblfloatfix}
\usepackage{placeins}

\def\BibTeX{{\rm B\kern-.05em{\sc i\kern-.025em b}\kern-.08em
    T\kern-.1667em\lower.7ex\hbox{E}\kern-.125emX}}
\begin{document}

\title{A Hybrid Approach to Malware Detection: Integrating Few-Shot Model-Agnostic Meta-Learning with Autoencoders}

\author{\IEEEauthorblockN{Emmanuela Andam\textsuperscript{1,2}, Yasir Abbas Zaidi\textsuperscript{1,2}, Abdelali Hadir\textsuperscript{3}, Emmanuel Grant\textsuperscript{2}, and Naima Kaabouch\textsuperscript{1,2}}
\IEEEauthorblockA{\textsuperscript{1}Artificial Intelligence Research (AIR) Center, University of North Dakota, Grand Forks, ND, USA\\
\textsuperscript{2}School of Electrical Engineering and Computer Science, University of North Dakota, Grand Forks, ND, USA\\
\textsuperscript{3}National School of Commerce and Management, Hassan II University, Casablanca, Morocco}}

\maketitle

\begin{center}
\footnotesize
\textcopyright~2025 IEEE. Personal use of this material is permitted. Permission from IEEE must be obtained for all other uses, in any current or future media, including reprinting/republishing this material for advertising or promotional purposes, creating new collective works, for resale or redistribution to servers or lists, or reuse of any copyrighted component of this work in other works.
\end{center}

\begin{abstract}
Ransomware has emerged as a major cybersecurity threat, with incidents increasing in frequency and impact across critical sectors. These attacks are typically launched through phishing emails, malicious downloads, or exploitation of software vulnerabilities to gain system access. Once inside, the malware encrypts files and demands a ransom, often in cryptocurrency, for the decryption key. Conventional detection methods often struggle with novel or scarce samples, leaving systems vulnerable. To address these challenges, this paper proposes a hybrid deep learning framework that combines an Autoencoder Feature Extractor (AFE) with a Model Agnostic Meta Learning (MAML) classifier for few shot malware detection. The AFE generates compact latent features that reduce noise and dimensionality, while the MAML classifier rapidly adapts to new threats using limited labeled data. Experiments conducted on the Ransomware Dataset 2024 demonstrate the effectiveness of the framework in binary classification tasks. Across one to fifty shot settings, the proposed model consistently achieves high accuracy, F1 score, and Matthews Correlation Coefficient values, maintaining reliable classification even under extreme scarcity. These results highlight the model's robustness and effectiveness in adapting to limited data scenarios, demonstrating the potential of combining feature extraction with meta learning to enhance resilience against malware, particularly in sectors such as healthcare, manufacturing, and public infrastructure, where cyberattacks can cause significant operational and financial disruption.
\end{abstract}

\begin{IEEEkeywords}
malware detection, model-agnostic meta-learning (MAML), autoencoder, few-shot learning, feature extraction.
\end{IEEEkeywords}

\section{Introduction}
The proliferation of connected devices in the Internet of Things (IoT) era has fundamentally reshaped interactions between individuals and digital systems. However, this rapid expansion has also intensified the risk of cyberattacks, with malware representing one of the most prevalent threats. Malware, or malicious software, encompasses any program intentionally developed to disrupt, damage, or exploit computer systems and networks. Numerous documented incidents underscore the significant risks and consequences associated with such attacks. For instance, the May 2024 Ascension Health ransomware attack, initiated by a successful phishing attack where an employee downloaded a malicious file, shut down electronic health record systems across multiple states, costing the health system approximately \$13 billion and contributing to a \$1.8 billion operating loss for fiscal year 2024 \cite{b1}. Likewise, in January 2025, the LOSTKEYS malware campaign targeted NGOs and institutions in developed nations using a phishing lure disguised as a CAPTCHA page, enabling credential theft and the exfiltration of sensitive documents \cite{b2}. These cases demonstrate the escalating costs and complexity of cyberattacks, reinforcing the need for robust and adaptive malware detection systems. 

\begin{figure}[htbp]
    \centering
    \includegraphics[width=\columnwidth]{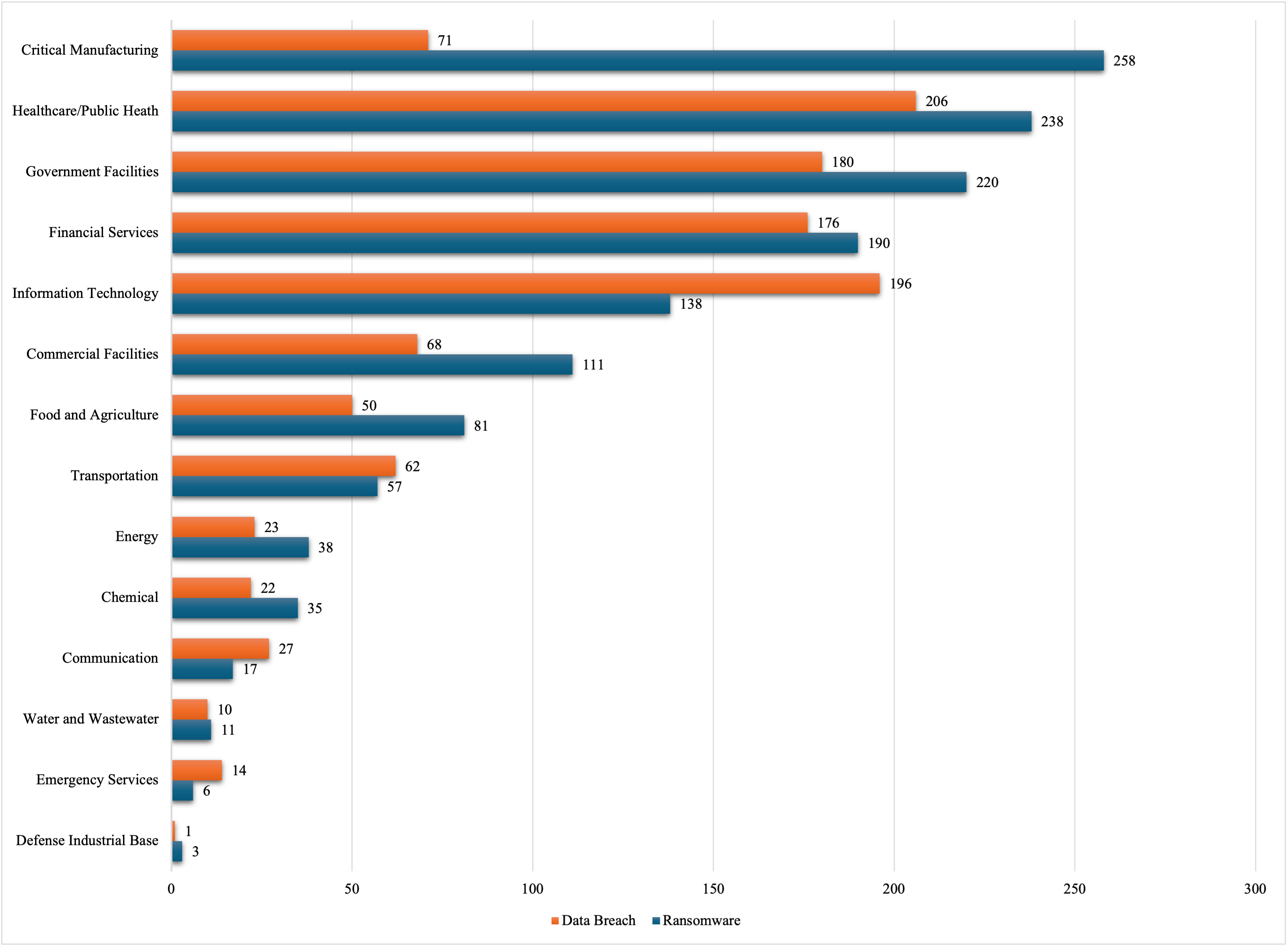}
    \caption{Reported Cyber Threats to Critical Infrastructure by Type and Sector, based on data from the FBI's 2024 Internet Crime Report}
    \label{fig:overview}
\end{figure}

The escalating financial toll of cybercrime is reflected in the sharp rise in reported losses, which increased from \$4.1 billion in 2020 to more than \$16.6 billion in 2024, a 33\% increase over the previous year. This financial burden is accompanied by a growing number of incidents targeting critical infrastructure \cite{b3}. Fig.~\ref{fig:overview} presents the number of reported ransomware attacks and data breaches across critical infrastructure sectors in 2024 \cite{b3}. The Healthcare and Public Health sector recorded the highest number of reported cyber incidents in 2024, including 238 ransomware attacks and 206 data breaches, while the Critical Manufacturing sector reported the greatest number of ransomware cases, totaling 258 \cite{b3}.

Numerous techniques have been proposed to detect malware attacks. Traditional methods like signature-based and heuristic approaches struggle to keep pace with malware detection \cite{b4}. Signature-based techniques rely on predefined patterns of known malware stored in databases \cite{b5}. Although effective against known threats, they are unable to detect newly modified or unseen malware variants \cite{b6}, \cite{b7}. Heuristic methods, which analyze code characteristics and execution patterns \cite{b8}, often produce high false positives in malware detection. Machine learning (ML) models such as Decision Trees (DT) and Support Vector Machines (SVM) are effective in improving the accuracy of detection while reducing false positives \cite{b8}. Though these techniques are effective, they are not efficient in detecting new malware attacks and require retraining on massive datasets.

One promising approach to handle these shortcomings is the combination of meta-learning and autoencoders.  This hybrid approach combines the strengths of both techniques to adaptability, generalization, and anomaly detection. 
\ Meta-learning (learning to learn) creates an environment where a model quickly adapts to new tasks by using prior knowledge, so it can detect new, unseen malware variants \cite{b9}. An autoencoder is an unsupervised neural network designed to learn compact representations of input data. It excels at modeling normal network behavior and flagging deviations indicating malicious activity. In this paper, we propose the integration of both architectures, Autoencoder Feature Extractor (AFE) with a Model-Agnostic Meta-Learning (MAML) classifier. This hybrid architecture employs the unsupervised feature learning capability of the autoencoder to create a compact, low-dimensional representation of the data. This compact representation is then used by the MAML classifier, which is optimized to quickly adapt to new few-shot tasks, enabling it to adapt dynamically to evolving malware patterns with minimal training data, thereby addressing the limitations of conventional ML-based models.

The remaining of this paper is structured as follows: Section II describes the methodology employed in this study, including the dataset and its preprocessing, model framework, and evaluation metrics used to compare the performance of the proposed model with that of others. Section III describes the findings, their significance, and limitations. Section IV concludes the study and discusses broader implications.

\section{Methodology}
In this section, the dataset and data-preprocessing, the proposed hybrid model, and evaluation metrics are discussed as follows:
\subsection{Dataset Collection and Preprocessing}
This study used the  Ransom Dataset 2024 from Zenodo.org \cite{b10}. This dataset comprises 21,752 samples, equally divided into 10,876 malicious and 10,876 benign samples, providing a balanced distribution suitable for machine learning tasks. It includes 11 well-known ransomware variants: Cerber, DarkSide, Dharma, GandCrab, LockBit, Maze, Phobos, REvil, Ragnar Locker, Ryuk, Shade, and WannaCry. These variants have been responsible for significant cyberattacks in recent years. The dataset captures 72 features describing processes, file activities, registry operations, network behavior, and API/DLL classifications, offering a comprehensive representation of malware behavior essential for effective detection. To prepare the dataset for modeling, preprocessing addresses the heterogeneity of numerical and categorical features. Categorical features, such as malware family labels, are transformed using label encoding, assigning unique integers to each category to enable numerical processing without introducing ordinal bias. Numerical features are scaled to the range [0, 1] via Min-Max normalization, defined as:
\begin{equation}
x_{\text{norm}} = \frac{x - x_{\min}}{x_{\max} - x_{\min}}\label{eq1}
\end{equation}
\begin{figure}[!b]
    \centering
    \includegraphics[width=\columnwidth]{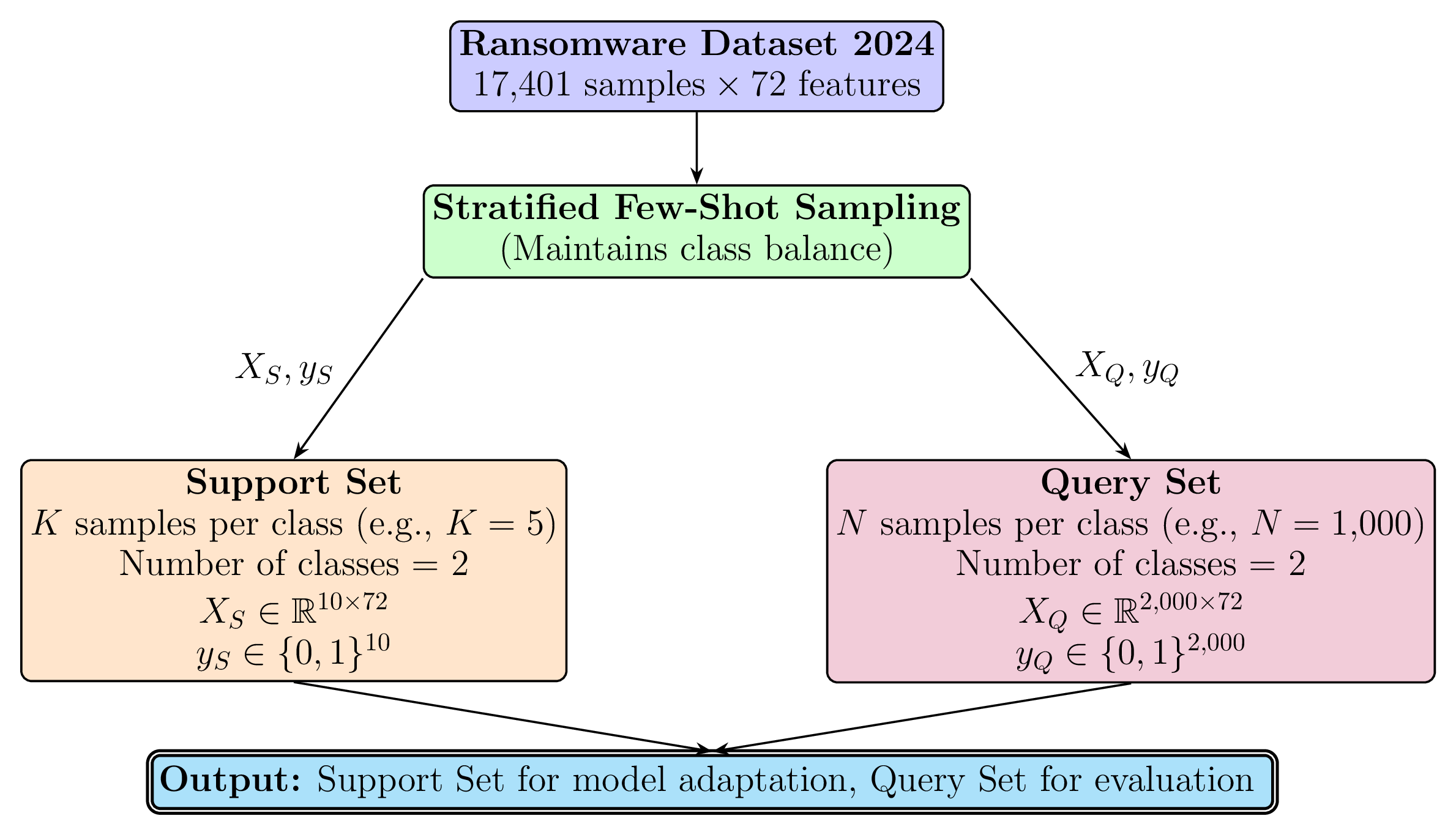}
    \caption{Few-shot sampling workflow showing dataset partitioning into support (K = 5 per class) and query (N = 2000 per class) sets for MAML training.}
    \label{fig:fewshot}
\end{figure}
\begin{figure*}[!b]
    \centering
    \includegraphics[width=\textwidth]{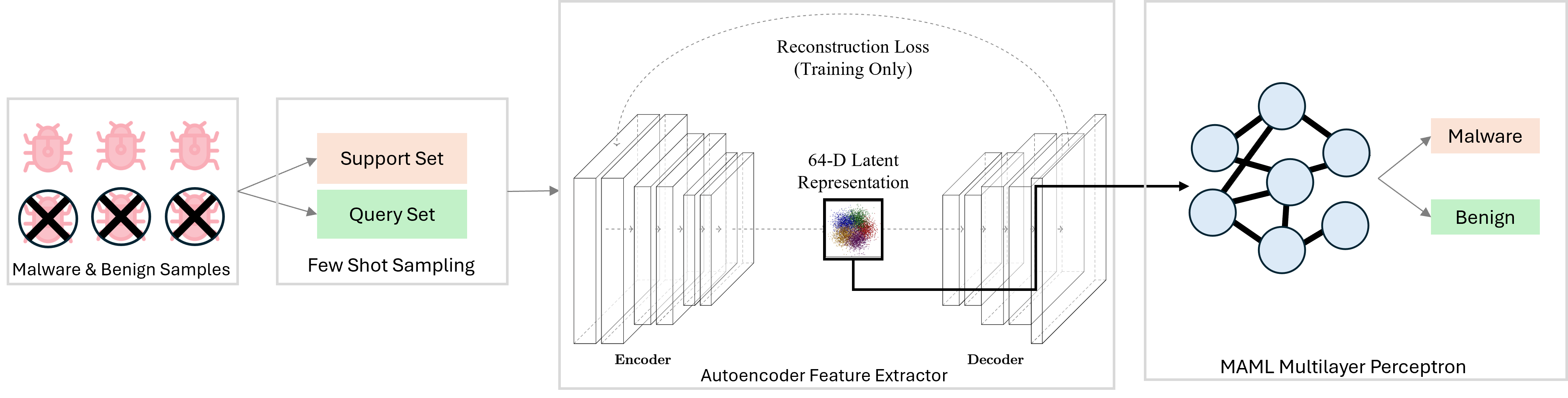} 
    \caption{Pipeline for few-shot malware detection. The process begins with stratified sampling to create support and query sets, followed by an autoencoder for feature extraction, and finally a classifier for malware detection.}
    \label{fig:picture2}
\end{figure*}
where $x_{\min}$ and $x_{\max}$ are the minimum and maximum values of each feature across the dataset.This normalization prevents high-magnitude features (e.g., API call frequencies) from dominating the learning process and accelerates model convergence.
To overcome the challenge of detecting new ransomware variants with limited labeled data, this study employs a few-shot learning paradigm. This approach, a component of meta-learning, trains a model to generalize to unseen classes by learning an effective adaptation strategy from a small number of examples. The methodology is structured around the standard N-way K-shot framework, where a new task is created in each training and evaluation episode. In our work, each task is a binary classification problem designed to differentiate between malware families and benign samples. Therefore, our model is configured for a 2-way K-shot task. This setup is a step toward evaluating the model's ability to handle novel threats, which is a requirement in a dynamic cybersecurity environment. For each episode, the dataset is divided into a support set and a query set as shown in Fig. ~\ref{fig:fewshot}. This procedure ensures that the model's training and evaluation accurately reflect the scarcity of labeled data for new threats. To mitigate potential biases from imbalanced classes in sampling, we employ stratified selection, ensuring equal representation.  
\begin{table}[!b]
\caption{Core Few-Shot Learning Parameters}
\label{table:fs_params}
\centering
\resizebox{\columnwidth}{!}{%
\begin{tabular}{lcl}
\hline
\textbf{Parameter} & \textbf{Value} & \textbf{Description} \\
\hline
$N$ (N-way) & 2 & Number of distinct classes per task (e.g., malware vs. benign) \\
$K$ (K-shot) & 5 & Number of labeled samples per class in the support set \\
$Q$ & 10{,}000 & Number of samples per class in the query set \\
\hline
\end{tabular}
}
\end{table}
\begin{table}[!t]
\caption{Architecture of the Autoencoder and the MAML Classifier}
\label{table:arch_details}
\centering
\resizebox{\columnwidth}{!}{%
\begin{tabular}{lcccl}
\hline
\textbf{Layer Name} & \textbf{Layer Type} & \textbf{Input Dim.} & \textbf{Output Dim.} & \textbf{Activation} \\
\hline
Encoder Layer 1 & nn.Linear & 72 & 256 & ReLU \\
Encoder Layer 2 & nn.Linear & 256 & 128 & ReLU \\
Encoder Layer 3 & nn.Linear & 128 & 64 & ReLU \\
Latent Space (Vector) & - & 64 & 64 & - \\
Classifier Layer 1 & nn.Linear & 64 & 128 & ReLU \\
Classifier Layer 2 & nn.Linear & 128 & 64 & ReLU \\
Classifier Layer 3 & nn.Linear & 64 & 2 & Softmax \\
\hline
\end{tabular}%
}
\end{table}
The specific parameters and their functions are detailed in the following sections:

\textit{Support Set} $(X_S, y_S)$: The support set is a limited collection of labeled samples, referred to as $K$-shots. This set is used for the inner-loop adaptation phase of the MAML algorithm, directly simulating the practical challenge of having only a handful of labeled samples for a newly emerging threat.

\textit{Query Set} $(X_Q, y_Q)$: The query set is composed of a larger number of samples per class, where $Q > K$. This set is used to evaluate the model's performance after it has adapted using the support set. The query set serves two primary functions: it computes the outer-loop loss during training and provides a reliable measure of the model's generalization capabilities during evaluation. The substantial size of the query set ensures that performance metrics are statistically robust and skewed by a small sample size. The core parameters defining the fundamental framework are presented in Table ~\ref{table:fs_params}. 

\subsection{Proposed Hybrid Model}
The proposed hybrid model was designed to address the challenges of few-shot learning in malware detection and classification. It integrates an Autoencoder Feature Extractor with a Model-Agnostic Meta-Learning (MAML) classifier to enable effective adaptation to new, unseen malware families using limited labeled samples. The autoencoder provides unsupervised feature learning to generate compact 64-dimensional latent representations from raw malware and benign samples, while the MAML component allows for rapid, task-specific fine-tuning during inference. As illustrated in Fig.~\ref{fig:picture2}, the complete methodology pipeline begins with stratified few-shot sampling to create balanced support and query sets, followed by autoencoder-based feature extraction, and culminates in MAML-driven classification for malware detection. The components of the model’s architecture (e.g., encoder/decoder layers) are detailed in Table ~\ref{table:arch_details}.
\begin{figure*}[!t]
    \centering
    \includegraphics[width=\textwidth]{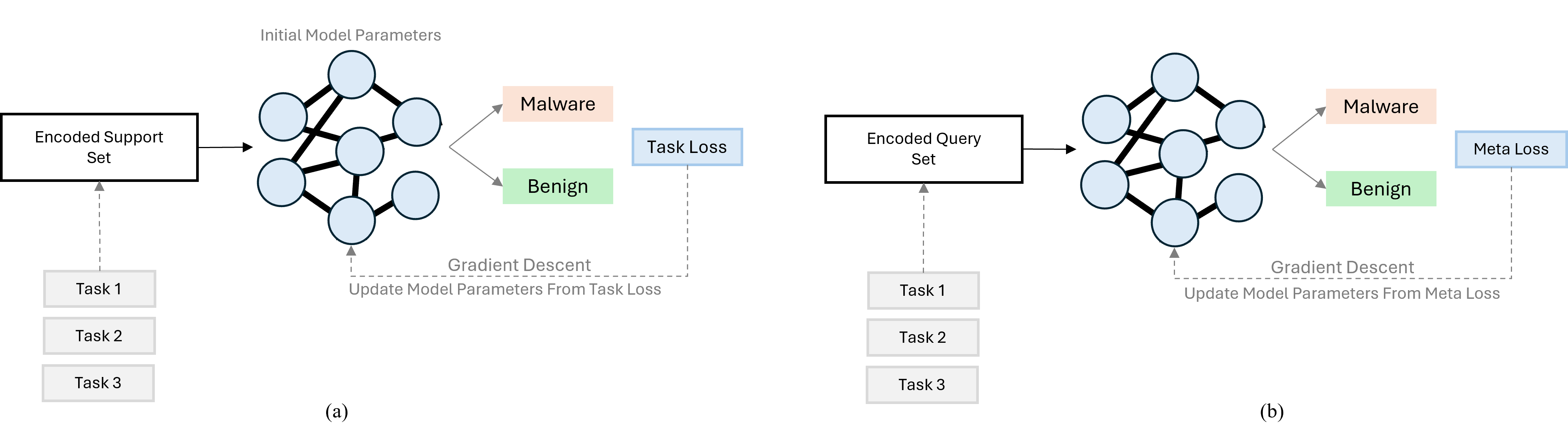} 
    \caption{Model-Agnostic Meta-Learning training process, showing (a) the inner-loop task adaptation where model parameters are updated based on a task's loss, and (b) the outer-loop meta-optimization where the original parameters are updated to improve future adaptation.}
    \label{fig:picture3}
\end{figure*}
\paragraph{Autoencoder Feature Extraction}
The AFE is an autoencoder, a neural network designed for unsupervised learning. Its architecture consists of an encoder and a decoder. The encoder maps high-dimensional input data to a compressed, latent representation, while the decoder attempts to reconstruct the original input from this representation. The AFE's encoder acts as a robust feature learning network, capturing the most salient features of malware samples in a compressed, low-dimensional latent space. This process reduces the risk of overfitting, which is a pervasive problem when dealing with small training sets.

\textit{Encoder}: This component is a multi-layer feedforward neural network that maps high-dimensional input features, $x \in \mathbb{R}^{72}$, to a lower-dimensional latent vector, $z \in \mathbb{R}^{64}$. The encoder architecture compresses the 72-dimensional input features to a 64-dimensional latent vector. The latent space dimension was empirically selected to optimize the trade-off between dimensionality reduction and information preservation.

\textit{Decoder}: This component mirrors the encoder's architecture, attempting to reconstruct the original input ($\hat{x}$) from the latent vector ($z$). The decoder reconstructs the 72-dimensional input from the 64-dimensional latent vector with similar hidden layers as the encoder.

The encoding and decoding process can be defined as
\begin{equation}
z = f_{\text{enc}}(x; \theta_{\text{enc}}), \quad 
\hat{x} = f_{\text{dec}}(z; \theta_{\text{dec}})
\end{equation}
where $\theta_{\text{enc}}$ and $\theta_{\text{dec}}$ represent the learnable parameters (weights and biases) of the encoder and decoder neural networks, respectively. These parameters are optimized during training via backpropagation to minimize reconstruction error.

The autoencoder is trained to minimize the reconstruction loss, which is the Mean Squared Error (MSE) between the input and the decoder’s output:
\begin{equation}
L_{AE} = \frac{1}{N} \sum_{i=1}^{N} \lVert x_i - \hat{x}_i \rVert^2 
\end{equation}

Minimizing this loss forces the encoder to capture the most salient features of the data, reducing noise and dimensionality, and improving its generalization capability. After the autoencoder training, the support and query sets are transformed as follows:
\begin{equation}
Z_S = f_{\text{enc}}(X_S), \quad Z_Q = f_{\text{enc}}(X_Q)
\end{equation}

\paragraph{Model-Agnostic Meta-Learning Classifier}
The latent features extracted from the autoencoder (central panel of Fig.~\ref{fig:picture2}) serve as inputs to the MAML classifier, which performs the final malware detection step (right panel of Fig.~\ref{fig:picture2}). To enable few-shot adaptation, the MAML algorithm employs a nested optimization loop. As illustrated in Fig.~\ref{fig:picture3}), this nested loop includes:

\textit{Inner Loop (Task Adaptation)}: For each few-shot task $\tau_i$ sampled from $p(\tau)$, a small number of gradient descent steps are taken on the support set. This process updates the model's parameters, $\theta$, to task-specific parameters, $\theta_i'$, using the loss on the support set $L_S$:  
\begin{equation}
\theta_i' = \theta - \alpha \nabla_\theta L_S(\theta)
\end{equation}
where $\alpha$ is the inner-loop learning rate.

\textit{Outer Loop (Meta-Optimization)}: The performance of these adapted parameters, $\theta_i'$, is then evaluated on the query set. The original parameters, $\theta$, are updated based on the performance of the adapted parameters to improve future adaptation. The outer-loop update is:  
\begin{equation}
\theta \leftarrow \theta - \beta \nabla_\theta \sum_i L_Q(\theta_i')
\end{equation}
where $\beta$ is the meta-learning rate.

This meta-gradient update learns an initialization that is sensitive to task-specific changes, allowing the model to adapt rapidly and generalize effectively to unseen malware categories
\paragraph{Evaluation Metrics}
In this work, we evaluated and compared the performance of the proposed model with those of other techniques in terms of metrics derived from the confusion matrix (comprising true positives (TP), true negatives (TN), false positives (FP), and false negatives (FN)). The metrics, along with their formulas and purposes, are detailed in Table~\ref{table:metrics}.

\section{Results and Discussion}
The experiments were conducted across various few-shot settings to simulate real-world conditions where labeled malware samples are limited and the model must adapt quickly to \begin{table}[!t]
\caption{Summary of Evaluation Metrics}
\label{table:metrics}
\centering
\renewcommand{\arraystretch}{1.4} 
\resizebox{\columnwidth}{!}{%
\begin{tabular}{l p{4.5cm} p{3.3cm}}
\hline
\textbf{Metric} & \textbf{Formula} & \textbf{Purpose} \\
\hline
Accuracy & $\displaystyle \frac{TP + TN}{TP + TN + FP + FN}$ & Proportion of correct predictions out of all predictions. \\
Precision & $\displaystyle \frac{TP}{TP + FP}$ & Accuracy of positive predictions. \\
Recall & $\displaystyle \frac{TP}{TP + FN}$ & Ability to detect all positive samples. \\
F1-Score & $\displaystyle 2 \cdot \frac{\text{Precision} \cdot \text{Recall}}{\text{Precision} + \text{Recall}}$ & Balance between precision and recall. \\
MCC & $\displaystyle 
\frac{TP \cdot TN - FP \cdot FN}
{\sqrt{
\makecell{
(TP + FP) \cdot (TP + FN) \\
\cdot (TN + FP) \cdot (TN + FN)
}}}
$ & Robust score even with imbalanced data. \\
\hline
\end{tabular}%
}
\end{table}
\begin{table}[!b]
\centering
\caption{Few-Shot Learning Parameters Used in Our Experiments}
\label{table:fewshot_params}
\renewcommand{\arraystretch}{1.3}
\begin{tabular}{c c c c c c c}
\hline
\textbf{K-shot} & 1 & 2 & 5 & 10 & 20 & 50 \\
\hline
\textbf{n\_query} & 2000 & 2000 & 2000 & 2000 & 2000 & 2000 \\
\hline
\end{tabular}
\end{table} evolving threats or new malware variants. The performance is assessed using the metrics presented in Table~\ref{table:metrics}. All experiments were carried out on a computer with a 12th Gen Intel(R) Core(TM) i7-12700 processor (2.10 GHz) and 32 GB of RAM, using the PyTorch framework. The model's key hyperparameters were set as follows: an inner-loop learning rate ($\alpha$) of 0.01, a meta-learning rate ($\beta$) of 0.001, and five inner-loop gradient steps.  

To thoroughly evaluate the model under varying data scarcity conditions, we experimented with a range of values for the key few-shot parameters, as summarized in Table~\ref{table:fewshot_params}. Specifically, we varied the number of shots ($K$-shot) from 1 to 50, while maintaining a fixed query size ($n_{\text{query}}$) of 1000 samples per class to construct balanced evaluation sets totaling 2000 samples (1000 benign and 1000 malware). This setup allows for consistent comparisons across shot levels, focusing on how additional support samples enhance the model's rapid adaptation capabilities in scenarios that mimic zero-day threats or newly emerging malware families.

\begin{figure}[!t]
\centering
\includegraphics[width=\columnwidth]{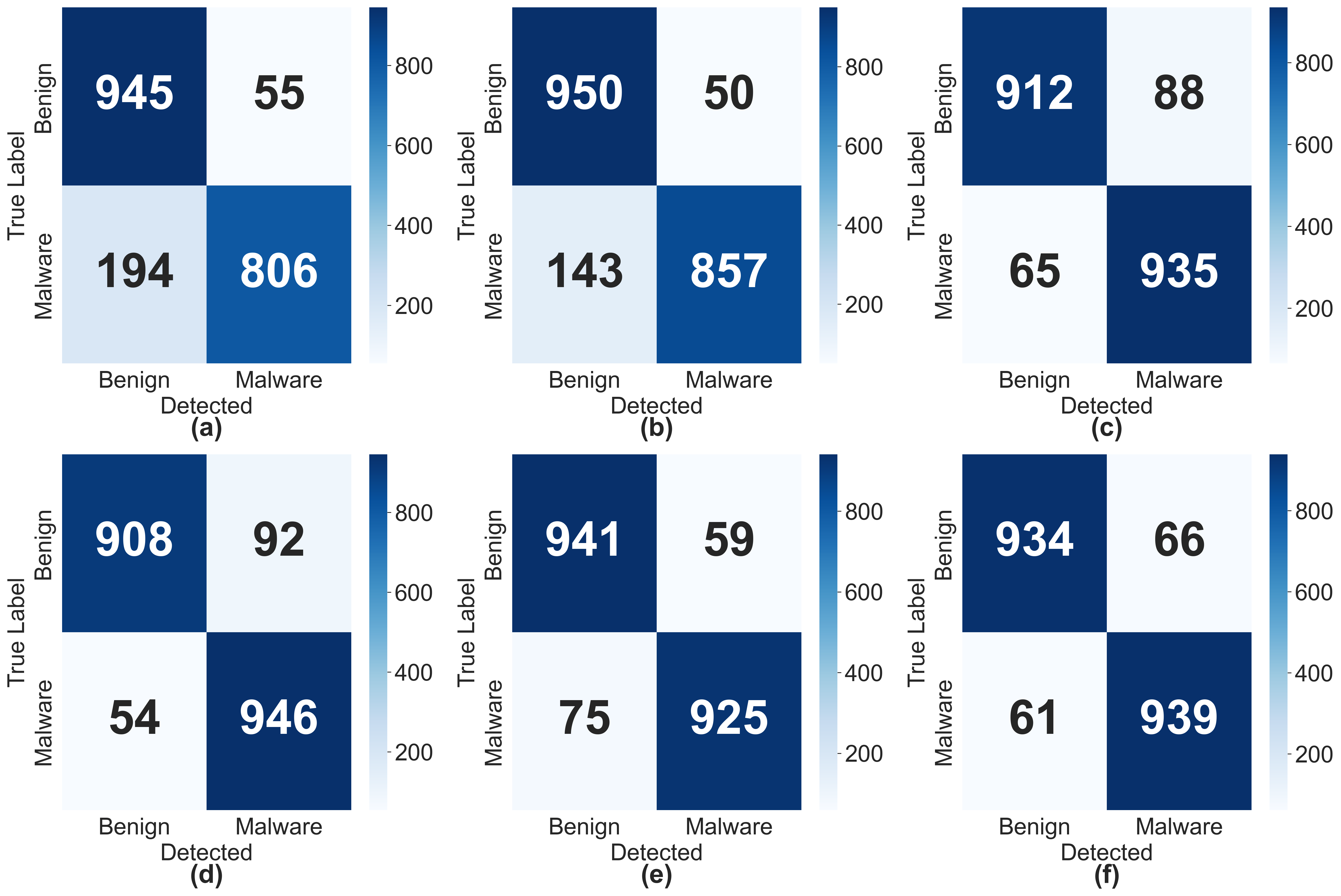}
\caption{Confusion matrices illustrating the model's classification performance in distinguishing benign from malware samples across few-shot settings: (a) 1-shot, (b) 2-shot, (c) 5-shot, (d) 10-shot, (e) 20-shot, and (f) 50-shot, each with a fixed query size of 1000 samples per class.}
\label{fig:confusion_matrices}
\end{figure}

Fig.~\ref{fig:confusion_matrices} shows the confusion matrices which illustrate the model's ability to distinguish malware from benign samples under varying levels of data scarcity ranging from 1-shot to 50-shot settings. In the 1-shot scenario , the model correctly identified 945 benign samples and 806 malware samples, but exhibited 55 false positives (benign misclassified as malware) and 194 false negatives (missed malware detections), reflecting challenges for extreme data size limitation where the model has minimal examples to learn from. As the number of shots increases, such as in the 2-shot setting with 950 true negatives and 857 true positives, the number of errors decreased, showing improved discrimination. This trend continues in higher shots; for instance, the 5-shot matrix achieved 912 true negatives and 935 true positives with only 88 false positives and 65 false negatives, while the 50-shot reached 934 true negatives and 939 true positives, minimizing errors to 66 false positives and 61 false negatives. Despite some persistent false negatives in lower-shot settings, potentially due to subtle variations in malware features, the high true positive rates across all matrices highlight the model's robust generalization, confirming the value of the MAML-based approach for evolving malware threats. These matrices underscore the practical  use in cybersecurity where minimizing missed detections is important to prevent breaches.

Building on these per-instance insights from the confusion matrices, the overall performance across all few-shot scenarios is summarized in Table~\ref{table:performance_fewshot}, \begin{table}[!b]
\centering
\caption{Performance Across Different Few-Shot Settings}
\label{table:performance_fewshot}
\renewcommand{\arraystretch}{1.3}
\resizebox{\columnwidth}{!}{%
\begin{tabular}{c c c c c c}
\hline
\textbf{Experimental Setting} & \textbf{Accuracy} & \textbf{F1 Score} & \textbf{Precision} & \textbf{Recall} & \textbf{MCC} \\
\hline
1-shot  & 0.8755 & 0.8662 & 0.9361 & 0.8060 & 0.7584 \\
2-shot  & 0.9035 & 0.8988 & 0.9449 & 0.8570 & 0.8105 \\
5-shot  & 0.9235 & 0.9244 & 0.9140 & 0.9350 & 0.8472 \\
10-shot & 0.9270 & 0.9284 & 0.9114 & 0.9460 & 0.8546 \\
20-shot & 0.9330 & 0.9325 & 0.9400 & 0.9250 & 0.8661 \\
50-shot & 0.9365 & 0.9367 & 0.9343 & 0.9390 & 0.8730 \\
\hline
\end{tabular}%
}
\end{table}
\begin{figure}[!b]
    \centering
    \includegraphics[width=\columnwidth]{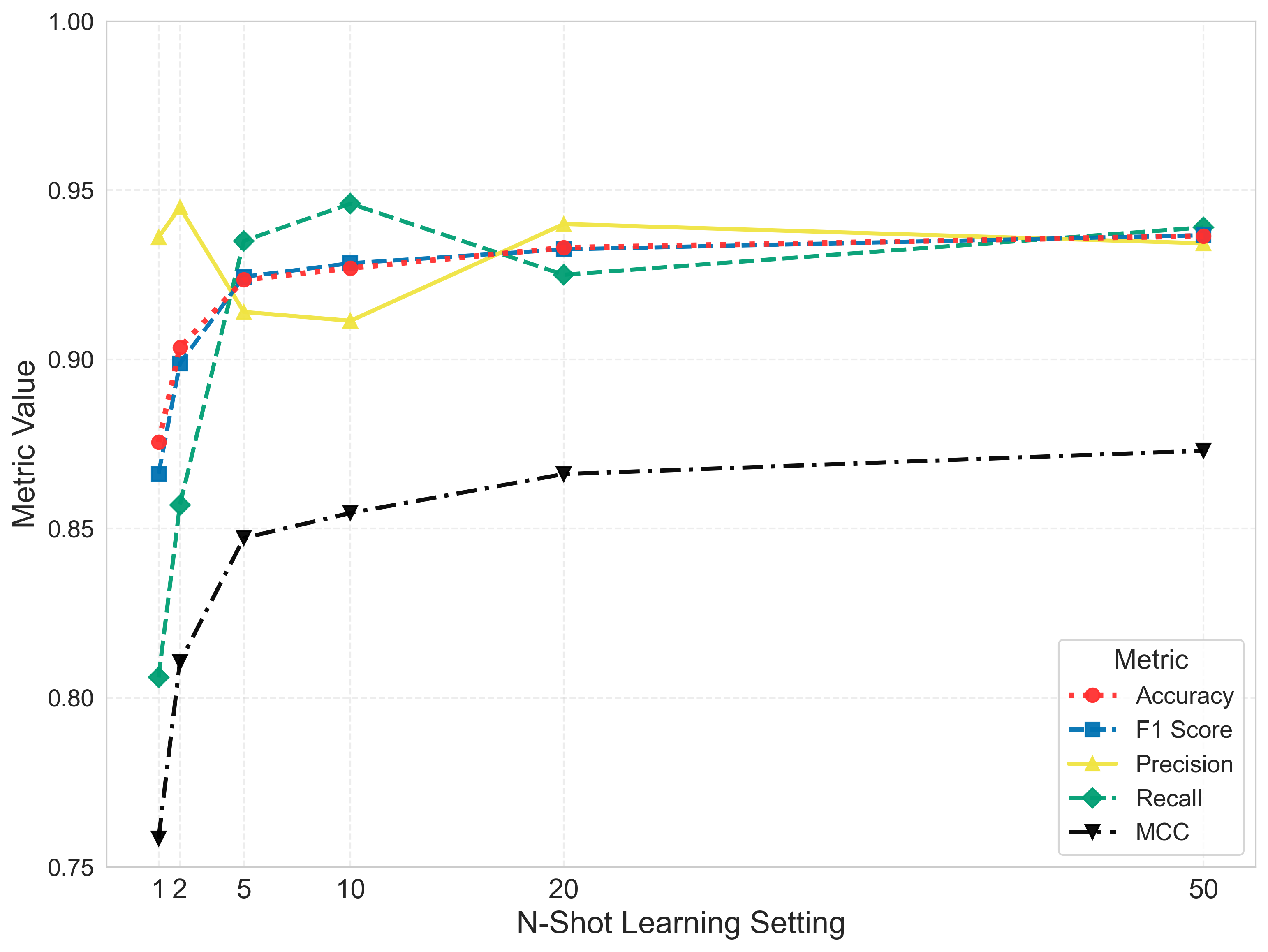}
    \caption{Line chart depicting the model's performance metrics (accuracy, F1 score, precision, recall, and MCC) across varying few-shot settings from 1-shot to 50-shot, with a fixed query size of 1000 samples per class. The upward trend in metrics highlights improved generalization and resilience to class imbalance with increasing support samples.}
    \label{fig:maml_performance}
\end{figure}
\begin{figure*}[!t]
    \centering
    \includegraphics[width=0.9\textwidth]{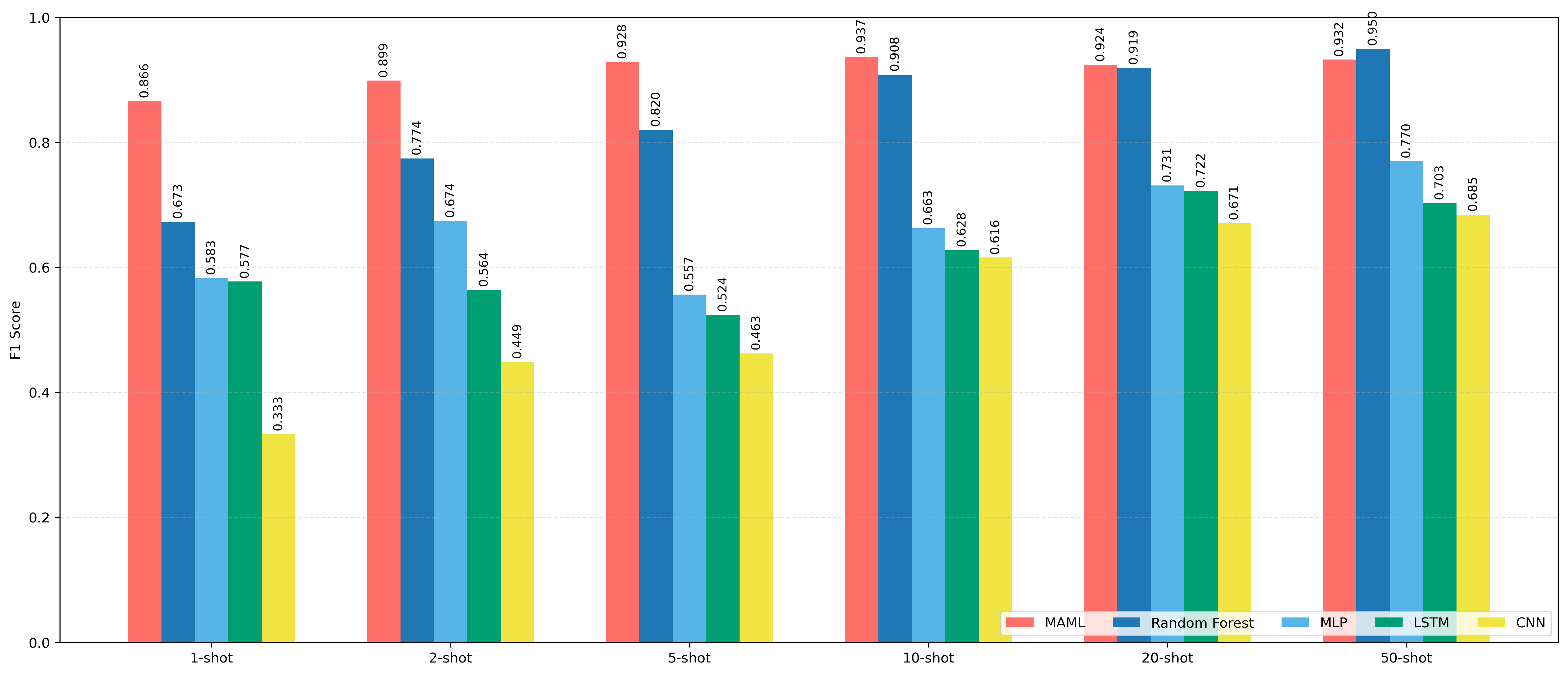}
    \caption{F1 score of AFE-MAML and baseline models across 1 to 50-shot settings.}
    \label{fig:maml_comparison_f1_score}
\end{figure*}
\begin{figure*}[!t]
    \centering
    \includegraphics[width=0.9\textwidth]{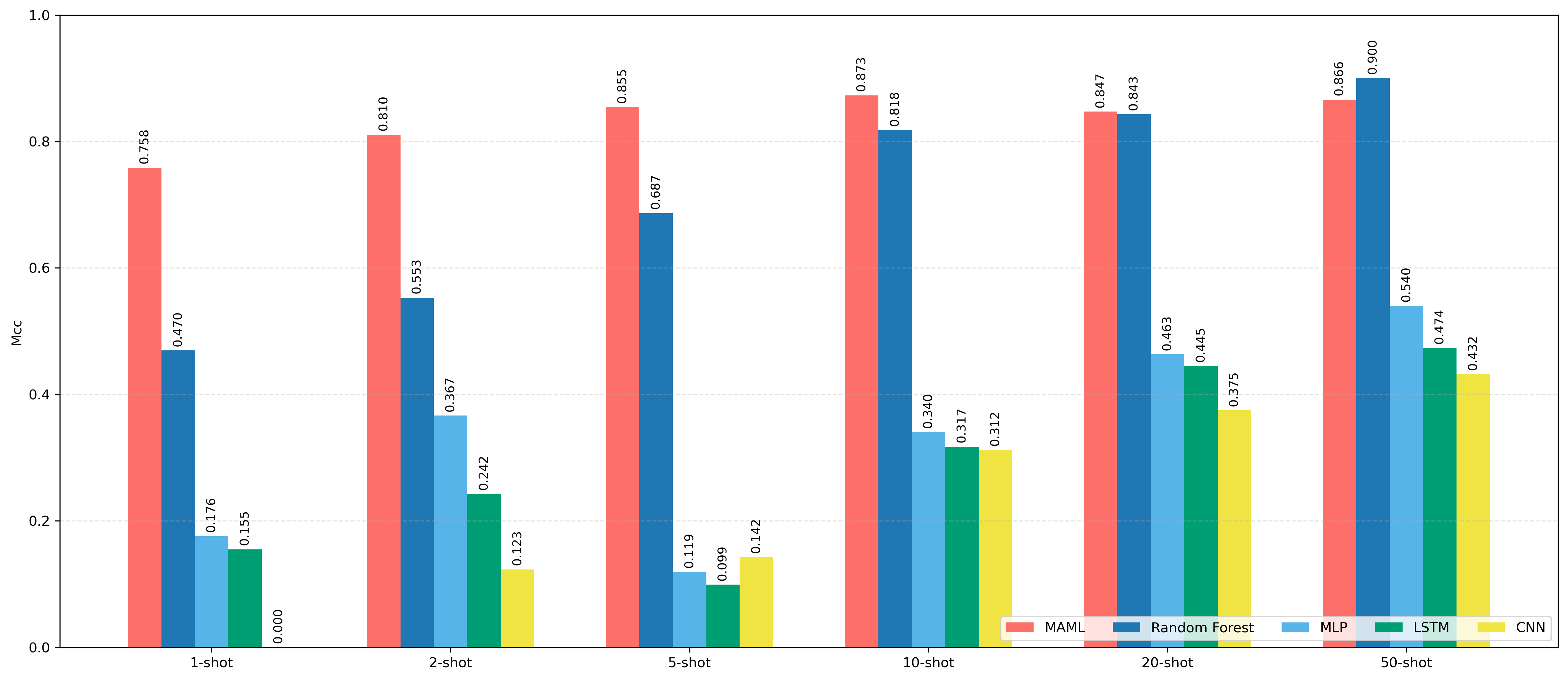}
    \caption{Matthews Correlation Coefficient of AFE-MAML and baseline models across 1 to 50-shot settings.}
    \label{fig:maml_comparison_mcc}
\end{figure*}
which aggregates key metrics (accuracy, F1 score, precision, recall, and MCC) for each configuration. These results demonstrate consistent, high performance, with the proposed model excelling in low-data regimes and showing progressive improvements as shot counts rise. For example, in the 1-shot setting, the model achieved an accuracy of 0.8755 and MCC of 0.7584, which, while solid for such limited data, improved markedly to 0.9365 accuracy and 0.8730 MCC in the 50-shot configuration. These aggregated metrics show that increased shots lead to better balance between precision (e.g., reducing false positives) and recall (e.g., detecting more true malware).

These results show that as the number the shots increases, the performance of the proposed model increases. For example, the 1-shot setting yielded an accuracy of 0.8755 and F1 of 0.8662 while the  for the 50-shot setting the model provided an accuracy of 0.9365, an F1 of 0.9367, and an MCC of 0.8730. The high F1-score and MCC values are particularly noteworthy as they confirm the model's resilience to the inherent class imbalance in malware datasets, where benign samples often exceed the number of malicious ones in real-world datasets. From a practical cybersecurity perspective, this is critical; a low F1-score could lead to a high rate of false negatives (missed malware detections), potentially allowing threats to propagate, while a low MCC could result in an unacceptable level of false positives (benign files flagged as malicious), leading to alert fatigue for security analysts and operational inefficiencies. The scores observed in lower-shot settings, such as the 5-shot F1 of 0.9244, are especially promising, demonstrating that for rapid, small-scale deployments, such as responding to novel ransomware or trojan variants, the model can achieve reliable detection without extensive retraining.

Fig. ~\ref{fig:maml_performance}  shows the values of 5 metrics (accuracy, F1-score, precision, recall, and MCC) as a function of the number of shots for the proposed model. As expected, most of these metrics increase with the number of shots. For instance, comparing the 5-shot (accuracy 0.9235) and 10-shot (0.9270) configurations, the performance with 10 shots is consistently higher across metrics (e.g., recall improves from 0.9350 to 0.9460), demonstrating that providing even a modestly larger support set within the few-shot paradigm allows the model to learn a more robust and generalizable adaptation.
\FloatBarrier
\begin{figure*}[!t]
    \centering
    \includegraphics[width=0.9\textwidth]{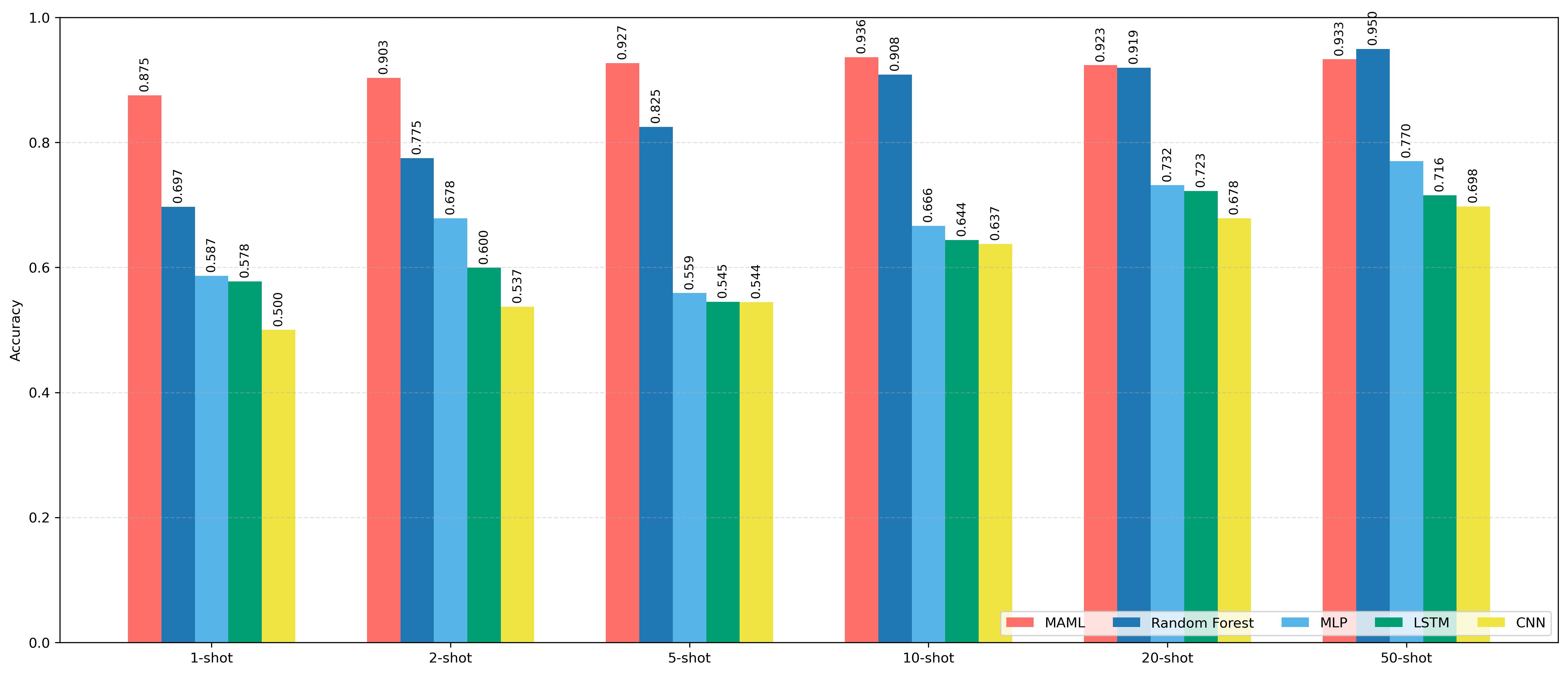}
    \caption{Accuracy of AFE-MAML and baseline models across 1 to 50-shot settings.}
    \label{fig:maml_comparison_accuracy}
\end{figure*}

Notably, precision shows a slight drop from 5-shot (0.9140) to 10-shot (0.9114) before recovering to 0.9400 at 20-shot; this temporary dip may arise as the model incorporates more diverse support samples, leading to minor over-adaptation and increased false positives in the intermediate range, before the decision boundaries stabilize with further data to balance precision and recall. This aligns with the confusion matrix observations, where higher shots correlate with reduced false negatives, enhancing recall without sacrificing precision. The strong performance in lower-shot regimes validates the efficacy of our few-shot learning approach, as it is designed for situations with extreme data scarcity, such as emerging threats in enterprise networks or mobile ecosystems where labeling new samples is costly and time-intensive.

Figs.~\ref{fig:maml_comparison_accuracy},~\ref{fig:maml_comparison_f1_score} and ~\ref{fig:maml_comparison_mcc} present the performance of the proposed MAML hybrid compared with baseline models (CNN, LSTM, MLP, and Random Forest) across different N-shot settings, in terms of the accuracy, F1 score, and the MCC to provide a comprehensive assessment of classification performance and robustness to class imbalance. All models were implemented under identical conditions to ensure a fair and accurate comparison.
Fig. ~\ref{fig:maml_comparison_accuracy} illustrates the accuracy of each model across N-shot settings. The MAML hybrid achieves consistently high accuracy, ranging from 0.876 at 1-shot to 0.933 at 50-shot, demonstrating its ability to adapt effectively in low-data scenarios. Random Forest shows steady improvement from 0.697 to 0.950, eventually surpassing the other three models in high-shot regimes. MLP and LSTM achieve moderate performance, ranging from 0.587 to 0.770 and 0.578 to 0.716, respectively, while CNN remains the weakest performer with a performance increasing from 0.500 to 0.698 across the N-shot spectrum. Fig. ~\ref{fig:maml_comparison_f1_score} shows the F1 scores, which emphasize the balance between precision and recall. MAML maintains superior performance, with scores of 0.866 at one-shot and 0.933 at fifty-shot, indicating reliable classification even under data scarcity. Random Forest follows a similar trend, improving from 0.673 to 0.950 with more shots. MLP and LSTM show moderate gains, from 0.583 to 0.770 and 0.577 to 0.703, respectively. Whereas CNN exhibits the lowest scores, starting at 0.333 and reaching 0.685. 

Fig. ~\ref{fig:maml_comparison_mcc} presents the Matthews Correlation Coefficient, which is particularly sensitive to class imbalance. MAML demonstrates strong correlation values ranging from 0.758 at one-shot to 0.866 at fifty-shot, with the peak at ten-shot (0.873) highlighting its classification quality. Random Forest shows improvement from 0.470 to 0.900, outperforming  the other baseline models in high-shot conditions. MLP and LSTM exhibit moderate performance, increasing from 0.176 to 0.540 and 0.155 to 0.474, respectively, while CNN remains the lowest at 0.000 to 0.432.

Overall, the proposed MAML hybrid is the best performer, consistently outperforming the baseline models in few-shot settings (1- to 20-shot) that align with the paper's emphasis on rapid adaptation to new threats with limited data. Additionally, our model outperforms other models proposed in the literature [11-13]. For instance, it surpasses the MI-MAML model, which yielded only 1 to 5\% accuracy improvements through data augmentation and adaptive learning rates \cite{b11}. Similarly, it exceeds the Relation-aware Siamese Denoising Autoencoder, which provided an accuracy between 83 and 97\% in 5-way 5-shot tasks and 90.3\% in obfuscated scenarios \cite{b12}. Furthermore, our model outperforms IoT attack detection frameworks, which achieved F1-scores of 0.75–0.90 in 5-shot cases \cite{b13}. The autoencoder's compact feature representations combined with MAML's rapid adaptability enable this enhanced generalization, laying the foundation for future obfuscation-resilient designs.

\section{Conclusion}
In this paper, we introduced a hybrid architecture that combines an Autoencoder Feature Extractor with a Model-Agnostic Meta-Learning classifier to address the problem of few-shot malware detection for emerging ransomware variants. This approach takes advantage of the autoencoder’s ability to create compact and informative feature representations which enable the MAML classifier to rapidly adapt to new, unseen threats with minimal labeled data. The results show  that the proposed model outperformed four traditional models in low-shot scenarios, achieving high accuracy, F1 score, and Matthews Correlation Coefficient values that reflect reliable classification even under class imbalance. While Random Forest becomes competitive in high-shot settings, the AFE-MAML hybrid maintains a clear advantage when labeled data are scarce, highlighting its suitability for rapid, small-scale deployment. The results indicate that this proposed framework provides a robust foundation for future research, including multi-class malware detection and obfuscation-resilient designs, contributing to the development of adaptive malware detection systems that are both accurate and flexible.

\vspace{12pt}

\end{document}